\documentclass[twocolumn,superscriptaddress,showpacs,aps]{revtex4}
\usepackage{makeidx}
\usepackage[dvips]{epsfig}
\usepackage{graphicx,color}
\newcounter{fig}

\begin{document}
\title{InN dielectric function from the midinfrared to the visible range}
\author{L.A. Falkovsky}
\affiliation{L.D. Landau Institute for Theoretical Physics, Moscow
117334, Russia} \affiliation{ Institute of the High Pressure
Physics, Troitsk 142190, Russia}
%\author{}

\date{\today}      % Deleting this command produces today's date.
\begin{abstract}
  The dispersion of the dielectric function for wurtzite InN
  is analytically evaluated in the  region near the fundamental energy gap.
  The real part of the dielectric function has a logarithmic singularity at the
absorption edge. This results in the large contribution into the
optical dielectric constant. For samples with degenerate carriers,
the real part of the dielectric function is divergent at the
absorption edge. The divergence is smeared with temperatures or
relaxation  rate. The imaginary part of the dielectric function
has a plateau far away from the absorption onset.
\end{abstract}
\maketitle

Recently, InN has attracted considerable attention due to its
potential application as other III-nitrides, but especially owing
to the small energy band gap $2\varepsilon_g$ of about 0.7 eV
observed \cite{Da,WWY,NSY,SSB,BF} in contrast to the value of 1.9
eV established for the last 20 years. The small band gap value
corresponds with a small effective electron mass  $m^{*}\approx
0,07m_0$  \cite{WWS,FC,RWQ}.

For future progress in  the research field, reliable material
parameters are derived from the most widespread {\it ab initio}
electronic-structure calculations. However, these methods do not
present  analytical results and  lead sometimes to contradictions
\cite{BFJ}, whether the 4d bands are included in the core or are
not. Therefore, the ${\bf k\cdot p}$ Hamiltonian is used to
clarify the physical content. In the corresponding Kane model
\cite{Ka} for the wurtzite case, the conduction-band and the
valence-band are constructed  from the $|s\rangle$ and
$|x\rangle$, $|y\rangle$, and $|z\rangle$ states at the
$\Gamma-$point.

\begin{figure}[]
\noindent\centering{
\includegraphics[width=80mm]{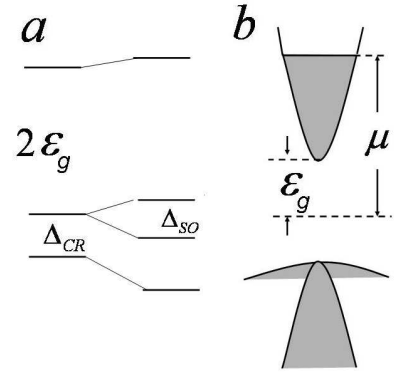}
} \caption{(a) scheme of the valence-band splitting under the
crystal field and the SO interactions; (b) the electron band, the
heavy-hole and light-hole bands near the
$\Gamma-$point.}\label{fig1}
\end{figure}

 In Fig. \ref{fig1}a, the scheme of the valence-band splitting in InN
 is shown under  the crystal field
$\Delta_{CR}$ and the spin-orbit interaction $\Delta_{SO}$.
According to experimental data \cite{GSW} and calculations
\cite{RWQ}, this splitting has a value on the order of $ 0.02\div
0.06$ eV, i. e. it is small in comparison with the band gap, and
can be ignored in calculations of the integral properties as the
optical absorption. Therefore,
 the Kane model can be substantially  simplified while using in
  calculations of the dielectric function (DF).

In this paper with the help of the simplified Kane model, we
evaluate analytically the DF for wurtzite InN in the range
$\omega\sim 0.3\div 4$ eV, where the absorption is dominated by
optical transitions from three highest valence bands into the
lowest conduction band. As well known, the imaginary part of the
DF has the square-root behavior near the absorption edge in the
case, if the conduction-band is empty  (i), and the step-like
behavior in the case, if the carriers in the conduction band are
degenerate  (ii). In the paper \cite{FHF}, the {\it ab initio}
calculations of the imaginary part were presented for the InN
polymorphs (with wurtzite, zinc-blend, and rocksalt structures) in
the case   (i), whereas the real part was restored using the
Kramers-Kronig relations. In the  paper \cite{FHF}, the imaginary
part was also estimated within  the Kane model, but these results
are misleading, because the optical-transition matrix elements are
incorrectly evaluated.

We find, that the real part of the DF  has
 at the absorption edge the kink-like singularity in the case (i)
 and the logarithmic divergence in the case (ii). Such singularities have been
previously obtained for graphene \cite{FV} and for IV-VI
semiconductors \cite{Fa}. The singularities are smeared with
temperatures or carrier relaxation. The excitonic effects was not
observed in InN, since they are suppressed by the carrier
relaxation (see \cite{WWS,FHF,TDH} for their estimation).

 The effective Hamiltonian of the simplified anisotropic Kane model
 is given as the matrix 4$\times$4:
\begin{equation}
 H =\left(
\begin{array}{cccc}
            \varepsilon_{g} & iP_2k_x          & iP_2k_y & iP_1k_z \\
          -iP_2k_x          & -\varepsilon_{g} & 0       & 0       \\
          -iP_2k_y          & 0                & - \varepsilon_{g} & 0 \\
          -iP_1k_z          & 0                & 0               & - \varepsilon_{g}
           \end{array}\right)\,,
\label{ham}\end{equation} where $P_1\ne P_2$ are the interband
momentum-matrix elements  with the velocity dimension (we put
 $\hbar=1$ in the intermediate formulas).
  Quadratic terms in the
momentum ${\bf k}$ can be written \cite{CC} in the main diagonal
of the matrix (\ref{ham}), as well as in the terms, connecting the
states at the valence band top,  $-\varepsilon_{g}$. We omit them,
because their contribution in the DF  is on the order of
$1/\ln{(\varepsilon_{at}/\varepsilon_{g})}\ll 1$, where
$\varepsilon_{at}$ is the energy of the atomic scale.

This  Hamiltonian (\ref{ham}) gives rise to the eigenvalues
\begin{equation}
\varepsilon_{1,4}=\pm[\varepsilon_g^2+P^2_{1}k_z^2+
P^2_2k_{\perp}^2]^{1/2},\quad k_{\perp}^2=k_x^2+k_y^2\,,
\label{eldis1}\end{equation}
 corresponding with the conduction band and the light-hole band,
 and the twofold eigenvalue
\begin{equation}\varepsilon_{2,3}=-\varepsilon_g \,,
\label{eldis2}\end{equation} for the heavy-hole band (see Fig.
\ref{fig1}b).

The effective masses at the conduction-band bottom are
\begin{equation}\label{em}
m_{\parallel}=\varepsilon_g/P_1^2\,, \quad
m_{\perp}=\varepsilon_g/P_2^2\end{equation} for the longitudinal
and transverse directions respectively to the main $z-$axis.
 Comparison with the values
\(m_{\parallel}=0.065 m_0,\, m_{\perp}=0.068 m_0,\)\, and
\(2\varepsilon_g=0.69 \) eV obtained \cite{RWQ} in the
experimental data analysis gives \(P_1=0.966\times10^8\) cm/sec,
\(P_2=0.945\times10^8\) cm/sec.

The velocity operator has the form
\[{\bf v}=\frac{\partial H}{\partial {\bf k}} =\left(
\begin{array}{cccc}
            0 & iP_2{\bf e}_x          & iP_2{\bf e}_y & iP_1{\bf e}_z \\
          -iP_2{\bf e}_x          & 0 & 0       & 0       \\
          -iP_2{\bf e}_y          & 0                & 0 & 0 \\
          -iP_1{\bf e}_z          & 0                & 0  & 0
           \end{array}\right)\]
%\end{equation}
with orthonormal vectors ${\bf e}_j$ chosen along the coordinate
axes.

Let us define the velocity matrix in the representation
diagonalizing the Hamiltonian (\ref{ham}):
\[{\tilde {\bf v}}=U^{\dagger}{\bf v}U.\]
Using the eigenfunctions of the Hamiltonian, one finds the matrix
$U$:
%\begin{equation}
%\begin{equation}
 \[U =\left(
\begin{array}{cccc}
 -a_1  & 0    & 0 & a_2 \\
 ip_x/ n_1    & -p_y/p_{\perp}    & p_xp_z/p_{\perp}p    & ip_x/n_2   \\
 ip_y/ n_1    & p_x/p_{\perp}   & p_yp_z/p_{\perp}p  & ip_y/ n_2  \\
 ip_z/ n_1    & 0          & -p_{\perp}/p & ip_z/n_2
           \end{array}\right)\,,\]
where we use the notations  \(p_z= P_1k_z\, ,
p_{x,y}=P_2k_{x,y}\,, p=\sqrt{p_z^2+p_{\perp}^2}\)\,,
\(n_{1,2}=\sqrt{2\varepsilon_1(\varepsilon_1\pm\varepsilon_g)}\)\,,
\(a_{1,2}=\sqrt{(\varepsilon_1\pm\varepsilon_g)/2\varepsilon_1}\),
and $\varepsilon_1$ is given in Eq. (\ref{eldis1}).

We note that the diagonal velocity-matrix elements in this
representation  coincide  with the derivative of the eigenvalues
\[\langle i|{\tilde {\bf v}}|i\rangle=\frac{\partial \varepsilon
_i}{\partial {\bf k}}\,\] and we write the off-diagonal elements
\begin{eqnarray}
\langle 1|{\tilde {\bf v}}|2\rangle
=i\sqrt{\frac{\varepsilon_1+\varepsilon_g}{2\varepsilon_1}}
P_2\frac{k_y{\bf e}_x-k_x{\bf e}_y}{k_{\perp}}\,,\qquad
\end{eqnarray}
\begin{eqnarray}\nonumber
\langle 1|{\tilde {\bf v}}|3\rangle =
\frac{-iP_1P_2}{\sqrt{2\varepsilon_1(\varepsilon_1-\varepsilon_g)}}
\left[\frac{k_z}{k_{\perp}}(k_x{\bf e}_x+k_y{\bf e}_y) -
k_{\perp}{\bf e}_z\right]\,,
\\\nonumber
\langle 1|{\tilde {\bf v}}|4\rangle
=\frac{\varepsilon_g}{\varepsilon_1\sqrt{\varepsilon_1^2-\varepsilon_g^2}}
(P_1^2k_z{\bf e}_z+P_2^2k_x{\bf e}_x+P_2^2k_y{\bf e}_y)\,,
\\
 \nonumber \langle 3|{\tilde {\bf
v}}|4\rangle=
\frac{-iP_1P_2}{\sqrt{2\varepsilon_1(\varepsilon_1+\varepsilon_g)}}
\left[\frac{k_z}{k_{\perp}}(k_x{\bf e}_x+k_y{\bf e}_y) -
k_{\perp}{\bf e}_z\right]\,.
\end{eqnarray}

These matrix elements enter the general quantum-mechanic formula
for the dynamic conductivity $\sigma_{\alpha \beta}(\omega)$
derived in the paper \cite{FV}.  Then, we obtain the DF with the
help of the relation
\begin{equation} \epsilon_{\alpha \beta}(\omega)=1+4\pi
i\sigma_{\alpha \beta}(\omega)/\omega\,. \label{df1}
\end{equation}
Due to the symmetry, the off-diagonal tensor components of the DF
vanish and there are only two independent components
$\epsilon_{zz}(\omega)$ and
$\epsilon_{xx}(\omega)=\epsilon_{yy}(\omega)$.

The DF is separated into the intraband and interband parts. The
intraband term contains only the diagonal velocity-matrix elements
and has the Drude-Boltzmann form. For instance, we obtain for the
degenerate electrons:
\begin{equation}\label{drud}
 \epsilon_{zz}^{intra}(\omega)= -\frac{4e^2P_1}{3\pi\hbar
P_2^2}\frac{(\mu^2-\varepsilon_g^2)^{3/2}}{\omega(\omega+i\nu)\mu}\,,
\end{equation}
where the chemical potential $\mu$, the relaxation rate $\nu$ and
the photon frequency $\omega$ are written in the common units.

Neglecting  the carrier relaxation, we can write the interband
term for the extraordinary component of the DF in the form
\begin{eqnarray}\label{con}\nonumber
\epsilon _{zz}^{inter}(\omega ) =1+ \frac{2e^2}{\pi^2} \int d^{3}k
\left\{\frac{ [f(-\varepsilon _{1})-f(\varepsilon_{1})]|{\tilde
v}_{14}^{z}|^2}{ 2\varepsilon
_{1}[4\varepsilon_{1}^2-(\omega+i\delta) ^{2}]}
\right.\\
\left.+\frac{ [f(-\varepsilon _{g})-f(\varepsilon_{1})]|{\tilde
v}_{13}^{z}|^2}{( \varepsilon
_{1}+\varepsilon_g)[(\varepsilon_{1}+\varepsilon_g)^2-(\omega+i\delta)^{2}]}
 \right.\\
\left.+\frac{ [f(-\varepsilon_{1})-f(-\varepsilon _{g})] |{\tilde
v }_{34}^{z}|^2}{( \varepsilon
_{1}-\varepsilon_g)[(\varepsilon_{1}-\varepsilon_g)^2-(\omega+i\delta)^{2}]}
 \right\}
 \,, \nonumber
\end{eqnarray}%
where  $f(\varepsilon)=1/(\exp{[(\varepsilon-\mu)/T]}-1) $ is the
Fermi function. For the pristine semiconductor at low temperature,
the conduction band is empty, but the chemical potential $\mu$ can
be higher  than the conduction band bottom $\varepsilon_g$ in the
case of doping (see Fig. \ref{fig1}b).

The different terms in the  braces present  the verious optical
transitions:  first, between the light-hole band and the
conduction band, second, between the heavy-hole band and the
conduction band, and  third, between the light- and heavy-hole
bands. The infinitesimal $\delta$ in the denominators of Eq.
(\ref{con}) defines the bypass around the poles. These bypasses
give the imaginary part of the DF, whereas the principal values of
the integrals yield the real part.

Transforming  the  integration variables
\begin{eqnarray}\nonumber
k_{x}=\frac{\sqrt{\varepsilon_1^2-\varepsilon_g^2}}{P_2}\sin{\theta}\cos{\varphi}\,,
k_{y}=\frac{\sqrt{\varepsilon_1^2-\varepsilon_g^2}}{P_2}\sin{\theta}\sin{\varphi}\,,\\
k_z=\frac{\sqrt{\varepsilon_1^2-\varepsilon_g^2}}{P_1}\cos{\theta}\,,
\frac{\partial(k_z,k_x,k_y)}{\partial(\varepsilon_1,\theta,\varphi)}=
\frac{\varepsilon_1\sqrt{\varepsilon_1^2-\varepsilon_g^2}}{P_1P_2^2}
\sin\theta\,,
\end{eqnarray}
we integrate over the angles $\theta$ and $\varphi$:
\begin{eqnarray}\label{con1}\nonumber
\epsilon _{zz}^{inter}(\omega ) = \frac{8e^2P_1}{3\pi \hbar P_2^2}
\int_{\varepsilon_g}^{\varepsilon_{at}} d\varepsilon
\sqrt{\varepsilon^2-\varepsilon_g^2}\left\{\frac{\varepsilon_g^2
[f(-\varepsilon)-f(\varepsilon)] }{ 2\varepsilon
^2[4\varepsilon^2-(\omega+i\delta) ^{2}]}
\right.\\
\left.+\frac{ f(-\varepsilon
_{g})-f(\varepsilon)}{(\varepsilon+\varepsilon_g)^2-(\omega+i\delta)^{2}}
 +\frac{f(-\varepsilon)-f(-\varepsilon _{g}) }{(\varepsilon-\varepsilon_g)^2
 -(\omega+i\delta)^{2}}
 \right\}+1
 \,. \nonumber
\end{eqnarray}%

The integral presenting the real part of the DF diverges
logarithmically at the upper limit. Since the leading contribution
arises from the values  $\varepsilon\sim (\mu, \omega)$, the
integral can be cut off at the atomic value of energy
$\varepsilon_{at}$, where our ${\bf k\cdot p}$ expansion becomes
incorrect.
 The imaginary part is easily evaluated for zero temperatures.
  For instance, we find for the case \(\mu>\varepsilon_g\),
  when  electrons fill the conduction band,
\begin{eqnarray}\label{con2}\nonumber
\text{Im}\,\epsilon _{zz}^{inter}(\omega )=\frac{4e^2P_1}{3\hbar
P_2^2}
 \left\{\frac{\varepsilon_g^2}{2\omega^2}
\sqrt{1-\frac{(2\varepsilon_g)^2}{\omega^2}}
\theta(\omega-2\mu)\right.
\\ +\left.\sqrt{1-\frac{2\varepsilon_g}{\omega}}
\theta(\omega-\varepsilon_g-\mu)
 %\\\left.+
 %\sqrt{1+2\varepsilon_g/\omega}[f(-\varepsilon_g-\omega)- f(-\varepsilon _{g}) ]
 \right\}
 \,,
\end{eqnarray}
where the step $\theta-$function conveys the condition for the
interband electron absorption.

Let us emphasize, that the band edge for the optical transitions
into the conduction band from the light-hole band at $\omega=
2\mu$  is higher than the edge for the transition from the
heavy-hole band at $\omega= \varepsilon_g+\mu.$ With increasing
the free electron concentration, both edges demonstrate the blue
Burstein-Moss shift.
% shown in Fig. \ref{fig3}.
At zero
temperatures, the chemical potential $\mu$, measured from the
midgap is determined by the free-electron concentration:
 \(n_0=(\mu^2-\varepsilon_g^2)^{3/2}/3 \pi^2\hbar^3 P_1P_2^2\,.\)
%\begin{figure}[b]
%\noindent\centering{
%\includegraphics[width=90mm]{bedge1.eps}
%} \caption{The blue shift of the absorption edge (in units of
% $2\varepsilon_g=0.69$ eV) for the optical transitions in the conduction band
% from the heavy-hole band (solid line) and from the light-hole (dotted line)
% with increasing of the free-electron concentration
%$n_0$.} \label{fig3}
%\end{figure}

 If the electrons are absent in the conduction band,
\(-\varepsilon_g<\mu<\varepsilon_g\)\,, the imaginary part of the
DF is given in  Eq.  (\ref{con2}) with substitution
\(\mu\rightarrow\varepsilon_g\,.\) Far away from the absorption
edges, where $\omega\gg \varepsilon_g+\mu$, the imaginary part
demonstrates the plateau-like character with
%In the plateau region \cite{GSC,KVM,GWC} 1-4.5 eV, 3.32
\begin{equation}\label{impa}\text{max\,Im}\,\epsilon ^{inter}(\omega
)=\frac{4e^2P_1}{3\hbar P_2^2}.\end{equation} The plateau noticed
also in the paper \cite{FHF}  and  for the  A$_4$B$_6$
semiconductors  in \cite{Fa} is a consequence of the linearity of
the electron dispersion at the energy larger in comparison with
the energy gap.

The real part of the DF contains the following contributions. The
transitions between the heavy-hole bands and the conduction band
give
\begin{eqnarray}\label{con3}\nonumber
 \text{Re}\,\epsilon_{zz}^{c,hh}(\omega)= \frac{8e^2P_1}{3\pi
\hbar P_2^2}\left\{
\ln{\frac{2\varepsilon_{at}}{\mu+\sqrt{\mu^2-\varepsilon_g^2}}}
\right.+ \frac{\sqrt{1+2x}}{2}\\ \times\left.
\ln{\frac{\mu+(\mu+\varepsilon_g)x+\sqrt{(\mu^2-\varepsilon_g^2)(1+2x)}}
{(\mu+\varepsilon_g+\omega)(1+x+\sqrt{1+2x})}}+F(x)
\right\}\end{eqnarray} where
 \begin{equation}\label{f}
F(x)=\frac{\sqrt{1-2x}}{2}
\ln{\frac{\mu-(\mu+\varepsilon_g)x+\sqrt{(\mu^2-\varepsilon_g^2)(1-2x)}}
{|\omega-\mu-\varepsilon_g|(1-x+\sqrt{1-2x})}}\,,
\end{equation}
if \(x=\varepsilon_g/\omega<1/2\)\, and
\[ F(x)=-\sqrt{2x-1}\arctan\frac{\omega\sqrt{2x-1}}
{\mu+\varepsilon_g-\omega+\sqrt{\mu^2-\varepsilon_g^2}}\,,\] if
\(x>1/2.\)
%\begin{eqnarray}\label{con3}\nonumber
% \text{Re}\,\epsilon_{zz}^{c,hh}(\omega)= \frac{4e^2P_1}{3\pi
%\hbar P_2^2}\left\{
%2\ln{\frac{2\varepsilon_{at}}{\mu+\sqrt{\mu^2-\varepsilon_g^2}}}
%\right.
%\\ +\sqrt{1-2x}
%\ln{\frac{\mu-(\mu+\varepsilon_g)x+\sqrt{(\mu^2-\varepsilon_g^2)(1-2x)}}
%{|\mu+\varepsilon_g-\omega|(1-x+\sqrt{1-2x})}}\\\nonumber\left.+
%\sqrt{1+2x}
%\ln{\frac{\mu+(\mu+\varepsilon_g)x+\sqrt{(\mu^2-\varepsilon_g^2)(1+2x)}}
%{(\mu+\varepsilon_g+\omega)(1+x+\sqrt{1+2x})}}
%\right\}\end{eqnarray} \(x=\varepsilon_g/\omega<1/2\)

% \begin{eqnarray}\label{con4}\nonumber
% \text{Re}\,\epsilon_{zz}^{c,hh}(\omega)= \frac{8e^2P_1}{3\pi
%\hbar P_2^2}\left\{
%\ln{\frac{2\varepsilon_{at}}{\mu+\sqrt{\mu^2-\varepsilon_g^2}}}
%+\frac{\sqrt{1+2x}}{2}\right.
%\\ \times \left.
%\ln{\frac{\mu+(\mu+\varepsilon_g)x+\sqrt{(\mu^2-\varepsilon_g^2)(1+2x)}}
%{(\mu+\varepsilon_g+\omega)(1+x+\sqrt{1+2x})}}\right.\\\nonumber
%\left.-\sqrt{2x-1}\arctan\frac{\omega\sqrt{2x-1}}
%{\mu+\varepsilon_g-\omega+\sqrt{\mu^2-\varepsilon_g^2}}\right\}\end{eqnarray}
%\(x=\varepsilon_g/\omega>1/2\)

The transitions between the light-hole band and the conduction
band contribute
 \begin{eqnarray}\label{con5}\nonumber
 \text{Re}\,\epsilon_{zz}^{c,lh}(\omega)= \frac{4e^2P_1x^2}{3\pi
 \hbar P_2^2}\left\{ 1-\sqrt{1-\varepsilon_g^2/\mu^2} \right.
\\\left. +\sqrt{1-4x^2}
\ln{\frac{\sqrt{\mu^2-\varepsilon_g^2}+\mu\sqrt{1-4x^2}}
{|(\omega/2)^2-\mu^2|^{1/2}(1+\sqrt{1-4x^2})}}\right\}\end{eqnarray}
for \(x<1/2\) and
\begin{eqnarray}\label{con6}\nonumber
 \text{Re}\,\epsilon_{zz}^{c,lh}(\omega)= \frac{4e^2P_1x^2}{3\pi
\hbar P_2^2}\left\{ 1-\sqrt{1-\varepsilon_g^2/\mu^2} \right.
\\\left. -\sqrt{4x^2-1}\arctan{\frac{(1-\sqrt{1-\varepsilon_g^2/\mu^2})
\sqrt{4x^2-1}}{\sqrt{1-\varepsilon_g^2/\mu^2}-1+4x^2}}\right\}
\end{eqnarray}
for \(x>1/2\,.\)

We find that the real part of the DF as a function of $\omega$
takes   at $x=1/2$ the maximal value for \(\mu=\varepsilon_g\):
\begin{eqnarray}\nonumber \text{max\,Re}\,\epsilon^{inter}_{zz}=1+
\frac{8e^2P_1}{3\pi\hbar
P_2^2}\\\times\left[\ln{\frac{2\varepsilon_{at}}{\varepsilon_g}}-
\frac{\ln{(3+2\sqrt{2})}}{\sqrt{2}}+\frac{1}{8}\right]\,.\label{kink}\end{eqnarray}
%7.206

If the carriers appear in the conduction band,
$\mu>\varepsilon_g$, the real part of the DF  has a logarithmic
singularity smeared with temperature or carrier relaxation. For
small relaxation rate $\nu$ in comparison with the photon
frequency $\omega$, we have in
 Eqs. (\ref{con2}), (\ref{f}),
(\ref{con5})  to substitute
\begin{eqnarray}\label{subs}
\theta(\omega-\omega_{at})\rightarrow\frac{1}{2}+\frac{1}{\pi}
\arctan[(\omega-\omega_{at})/2\nu]\\
\nonumber
(\omega-\omega_{at})^2\rightarrow(\omega-\omega_{at})^2+(2\nu)^2\,,
\end{eqnarray} where $\omega_{at}$ is the absorption
edge equal to  $ \varepsilon_g+\mu$ or $2\mu$ for the
corresponding transitions. If temperature plays a more important
role,  we should put $T$ instead of   $\nu$ in Eq. (\ref{subs}).

So far the extraordinary component $\epsilon_{zz}$ was presented.
The ordinary component $\epsilon_{xx}$ differs only in the factor
$P_2/P_1$, which equals   0.98 for the experimental values of the
effective masses  (\ref{em}).

Now the value of the cutoff parameter $\varepsilon_{at}$ is only
needed to calculate the DF. To estimate this value, we can use the
energy arising in the Kane model while the quadratic terms are
taken into account. According to estimations \cite{FHF,RWQ}, this
energy ranges from 8 to 15 eV. We take the intermediate value
$\varepsilon_{at}=10$ eV plotting
\begin{figure}[]
\noindent\centering{
\includegraphics[width=90mm]{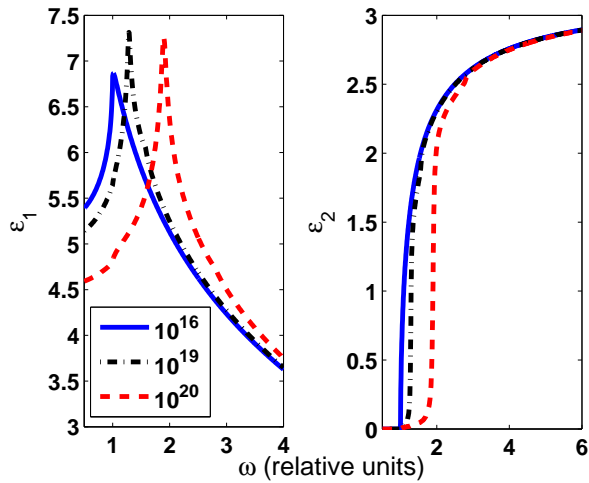}
} \caption{ The real and imaginary parts of the DF versus the
photon frequency (in units of the gap $2\varepsilon_g=0.69 $ eV)\,
for various free-electron concentrations, corresponding values of
the chemical potential $\mu$ are 1.01, 1.57, 2.79 (in units of
 $2\varepsilon_g=0.69$ eV); relaxation rate $\nu=$ 0,01
$\mu$.}\label{fig2}
\end{figure}
 Fig.  \ref{fig2}, where our theoretical results are shown. The maximum
value of the real part, Eq. (\ref{kink}), is found to equal 6.91
and the imaginary part of the DF   takes the value 3.16 on the
plateau, Eq. (\ref{impa}). The corresponding  values, $2.5\div
3.5$, obtained from experiments
  \cite{GSC,KVM,GWC} and  calculated from the first principles
  \cite{FHF}
   are on the plateau in the frequency range $1.5\div 4.0$ eV.
   That agrees very well with Fig.  \ref{fig2} (right panel).
   The experiment \cite{KVM} finds
   the  value about $\simeq 9$ for  the  maximum of the real part.
   The estimation \cite{KVM}
   of the dielectric constant gives  $\epsilon_{\infty}=6.7$.
   The  {\it ab initio} calculations
   \cite{CG, FHF} find correspondingly in these two papers
  $\epsilon_{\infty}(xx)=7.03$ and 7.16, as well as
  $\epsilon_{\infty}(zz)=7.41$ and 7.27. The agreement with our Fig.
  \ref{fig2} (left panel)
  is excellent  again. In our calculations, the maximum of the real part
  for the large carrier concentration increases logarithmically with decreasing of
  the relaxation rate. Plotting  Fig. \ref{fig2}, we take
$\nu=0.01\mu$ for various carrier concentrations.

  Concerning the dielectric constant $\epsilon_{\infty}$, we keep in
  mind  that the phonons contribute into its value. This
  contribution  can be estimated as
   $\omega_{TO}^2/\omega^2,$
 where  $\omega_{TO}$ is the transverse phonon frequency which is much
 less
 than the photon frequency considered here. Therefore, the phonon
 contribution into
 $\epsilon_{\infty}$ should be considered as negligible.

% Подчеркнем в заключение, что полученные здесь в явной
% форме аналитические результаты позволяют выяснить происхождение
% наблюдаемых особенностей диэлектрической проницаемости --
% большое значение и появление кинка у вещественной части, а также
% возникновение двух порогов поглощения при росте концентрации
% носителей и существование плато в поглощении выше порогов.

{\it In conclusions,} we find analytically that the real part of
the DF contains a singular contribution from the interband optical
transitions. It presents the large logarithmic term to the optical
dielectric constant. While increasing the frequency, we obtain the
dispersion of the dielectric function. Near the edge of the
interband absorption, a peak appears  in the real part of the DF
for degenerate electrons filled the conduction band if the
relaxation rate is large enough.

This work was supported by the Russian Foundation for Basic
Research (grant No. 07-02-00571). The author is grateful to the
Max Planck Institute for the Physics of Complex Systems for
hospitality in Dresden.

\end{document}